\documentclass{PoS}

\newcommand{\SU}{\mathrm{SU}}
\newcommand{\eq}[1]{\begin{equation}\label{#1}}
\newcommand{\en}{\end{equation}}
\newcommand{\ear}[1]{\begin{eqnarray}\label{#1}}
\newcommand{\enar}{\end{eqnarray}}

\title{Thermodynamics of the strongly interacting gluon plasma in the large-N limit}

\ShortTitle{Thermodynamics of the strongly interacting gluon plasma in the large-N limit}

\author{\speaker{M. Panero}\\%
        Institute for Theoretical Physics, ETH Zurich\\
        E-mail: \email{panero@phys.ethz.ch}}

\abstract{We report on our recent study of equilibrium thermodynamic observables in $\SU(N)$ gauge theories with $N=3$, $4$, $5$, $6$ and $8$ colors at temperatures $T$ in the range from $0.8 T_c$ to $3.4 T_c$ (where $T_c$ denotes the critical deconfinement temperature). The results, which show a very weak dependence on the number of colors, are compared with gauge/gravity models of the QCD plasma, including the improved holographic QCD model proposed by Kiritsis and collaborators, and with the supergravity prediction for the entropy density deficit. Furthermore, we investigate the possibility that the trace anomaly may receive contributions proportional to $T^2$ at temperatures close to $T_c$. Finally, we present the extrapolated results for the pressure, trace anomaly, energy and entropy densities in the $N \to \infty$ limit.}

\FullConference{The XXVII International Symposium on Lattice Field Theory\\
		 July 26-31, 2009\\
		 Peking University, Beijing, China}

\begin{document}

\section{Introduction and motivation}
\label{sec:intro}

Although several unquenched simulations of the QCD equation of state have appeared in recent years~\cite{lattice_thermodynamics_reviews}, lattice investigations of the thermodynamics of the Yang-Mills (YM) sector are still of interest from a fundamental point of view. In particular, the gluon sector is relevant for the limit in which the number of colors $N$ tends to infinity, where analytical simplifications take place~\cite{'tHooft:1973jz}. Furthermore, the large-$N$ limit is also a key-ingredient in the AdS/CFT conjecture~\cite{Maldacena:1997re} and in models of QCD based on gauge-gravity dualities~\cite{AdSQCD_reviews}. These motivations led me to address a numerical lattice study of the equilibrium thermodynamics properties at finite temperature in $\SU(N)$ gauge theories~\cite{Panero:2009tv}; similar works include refs.~\cite{similar_works}.\\
In this contribution, we discuss some of the findings of ref.~\cite{Panero:2009tv}, including, in particular, a comparison with the improved holographic QCD (IHQCD) model recently proposed by Kiritsis and collaborators~\cite{IHQCD_works} (see also refs.~\cite{related_works} for related works). The IHQCD model is an AdS/QCD model based on a 5D Einstein-dilaton gravity theory with a particular \emph{Ansatz} for the dilaton potential, which reproduces the leading terms of the $\SU(N)$ $\beta$-function at high energies, and linear confinement with a gapped, discrete glueball spectrum at low energies. The IHQCD model involves two free parameters, which in refs.~\cite{IHQCD_works} were fitted to match the results of previous lattice calculations.\\
In the context of gauge/gravity dualities, we also discuss the deficit of the entropy density $s$ with respect to its Stefan-Boltzmann limit $s_0$, and a possible comparison with the AdS/CFT prediction for the large-$N$ limit of the $\mathcal{N}=4$ supersymmetric YM theory~\cite{Gubser:1998nz}:
\eq{sugra_entropy}
\frac{s}{s_0} = \frac{3}{4} + \frac{45}{32} \zeta(3) ( 2 \lambda )^{-3/2} + \dots 
\en
in a temperature regime where the Yang-Mills plasma, while still strongly interacting, approaches approximate scale invariance.\\
Next, we investigate whether, in the temperature range studied, the trace anomaly of the deconfined $\SU(N)$ plasma $\Delta$ may receive contributions proportional to $T^2$~\cite{T2_papers}:
\eq{T2_contributions_to_Delta}
\frac{\Delta}{T^4} \stackrel{?}{=} \frac{A}{T^2}+B.
\en
Finally, we present an extrapolation of our results for $p/(N^2 T^4)$, $\Delta/(N^2 T^4)$, $\epsilon/(N^2 T^4)$ and $s/(N^2 T^3)$ to the $N \rightarrow \infty$ limit.

\section{Lattice setup}
\label{sec:lattice_setup}

We ran lattice simulations of $\SU(N)$ gauge theories with $N=3$, $4$, $5$ $6$ and $8$ colors on isotropic hypercubic lattices of sizes $N_s^4$ and $N_s^3 \times N_t$, with $N_s=20$ ($16$) for $N=3$ ($N>3$) and $N_t=5$. We used the standard, unimproved Wilson gauge action, and updated the configurations using heat-bath for $\SU(2)$ subgroups~\cite{heat_bath} and full-$\SU(N)$ overrelaxation~\cite{SUN_overrelaxation}. For $\SU(3)$, the physical scale was set using Sommer's parameter $r_0$~\cite{Necco:2001xg} while for $\SU(N>3)$ it was set by interpolating the string tension values taken from refs.~\cite{SUN_string_tensions}, or (at the largest $\beta$-values only) using the method of ref.~\cite{Allton:2008ty}.\\
On the lattice, the trace anomaly $\Delta = \epsilon - 3p$ is proportional to the difference of the expectation values of the plaquette at $T=0$ and at finite $T$:
\eq{trace}
\Delta  = T^5 \frac{\partial}{\partial T} \frac{p}{T^4} = \frac{6}{a^4} \frac{\partial \beta}{\partial \log a} \left( \langle U_\Box \rangle_{0} - \langle U_\Box \rangle_{T} \right).
\en
The pressure was determined using the ``integral method''~\cite{Engels:1990vr}:
\eq{pressure}
p  = T \frac{\partial}{\partial V} \log \mathcal{Z} \simeq \frac{T}{V} \log \mathcal{Z} = \frac{1}{a^4 N_s^3 N_\tau} \int_{\beta_0}^\beta d \beta^\prime \frac{\partial \log \mathcal{Z}}{\partial \beta^\prime}
= \frac{6}{a^4} \int_{\beta_0}^\beta d \beta^\prime \left( \langle U_\Box \rangle_{T} - \langle U_\Box \rangle_{0} \right),
\en
according to the integration methods discussed in ref.~\cite{Caselle:2007yc}. Eq.~(\ref{pressure}) can be affected by finite-volume corrections~\cite{Gliozzi:2007jh} (especially for very high temperatures~\cite{Endrodi:2007tq} or small volumes~\cite{small_volume_works}), but at the temperatures $0.8T_c \le T \le 3.4T_c$ investigated here, these effects are screened~\cite{Gross:1980br} and negligible within the data precision~\cite{Panero:2008mg}. The energy ($\epsilon$) and entropy ($s$) densities were obtained as: $\epsilon=\Delta+3p$, $s=(\Delta+4p)/T$.

\section{Results}
\label{sec:results}

\begin{figure}[-t]
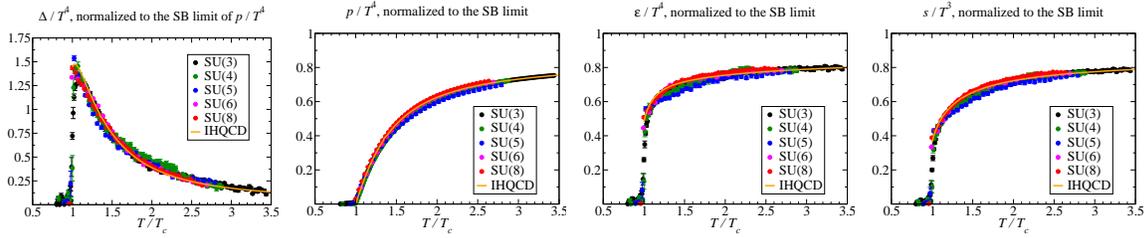

  \centerline{\includegraphics[width=.24\textwidth]{rescaled_trace.eps} \hfill \includegraphics[width=.24\textwidth]{pressure.eps} \hfill \includegraphics[width=.24\textwidth]{energy.eps} \hfill \includegraphics[width=.24\textwidth]{entropy.eps}}
  \caption{Comparison between the curves obtained using the improved holographic QCD model (yellow solid lines) and our lattice results for the trace anomaly, pressure, energy density and entropy density evaluated in the various $\SU(N)$ gauge groups. All quantities are normalized to their Stefan-Boltzmann (SB) limits, except for $\Delta/T^4$, which is normalized to the SB limit of $p/T^4$.} \label{fig:comparison_with_IHQCD}
\end{figure}

Fig.~\ref{fig:comparison_with_IHQCD} shows that the results for the rescaled trace, pressure, energy density and entropy density are very similar for all groups studied in this work\footnote{This feature would be compatible with some quasiparticle model descriptions~\cite{quasiparticle}.}; this makes it plausible that the QCD plasma could be described by models based on the large-$N$ limit. Fig.~\ref{fig:comparison_with_IHQCD} also shows the comparison with the improved holographic QCD model discussed above: the solid lines denote the results obtained in ref.~\cite{IHQCD_works}; the agreement between our $\SU(N)$ results and the IHQCD model is very good.

\begin{figure}[-t]
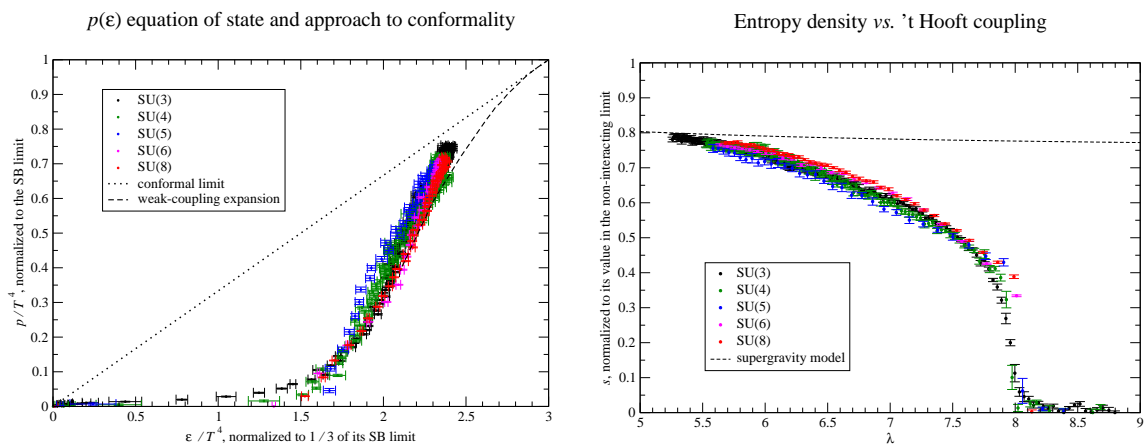

  \centerline{\includegraphics[width=.48\textwidth]{pressure_versus_energy.eps} \hfill \includegraphics[width=.48\textwidth]{entropy_lambda.eps}}
\vspace{10mm}

  \caption{Left panel: the equation of state, displayed as  $p(\epsilon)$, reveals the strong deviations of the $\SU(N)$ plasma from the dotted straight line corresponding to conformally invariant models, which is approached only at temperatures around $3T_c$; the dashed line is the weak-coupling expansion for $\SU(3)$ taken from ref.~\protect\cite{Hietanen:2008tv}. Right panel: the entropy density (normalized to its value in the free limit) as a function of the running 't~Hooft coupling in the $\overline{\rm MS}$ scheme; the dashed line is the corresponding prediction for strongly coupled $\mathcal{N}=4$ SYM~\protect\cite{Gubser:1998nz}, identifying the $\lambda$ parameter with the renormalized YM coupling.} \label{fig:AdSCFT_comparison}
\end{figure}

On the other hand, comparing the $\SU(N)$ simulation results with predictions derived from the $\mathcal{N}=4$ SYM model, is less straightforward: the regime in which the QCD plasma is strongly coupled is far from conformality---see the left panel of fig.~\ref{fig:AdSCFT_comparison}. In our simulations the $\SU(N)$ plasma approaches approximate scale invariance only at temperatures about $3T_c$; at such temperatures, it is still far from the Stefan-Boltzmann limit (top right corner of the diagram), and still strongly interacting. Interestingly, in that regime the $\SU(N)$ and the $\mathcal{N}=4$ SYM entropy densities (both normalized to their values in the free limit) appear to be close to each other, if one uses the value of the renormalized 't~Hooft coupling in the $\overline{\rm MS}$ scheme as the $\lambda$ parameter of the supersymmetric model (right panel of fig.~\ref{fig:AdSCFT_comparison}). Incidentally, we note that a comparison of $\mathcal{N}=4$ SYM and full-QCD lattice results for the drag force also yields $\lambda \simeq 5.5$~\cite{Gubser:2006qh}. Similar observations can be useful to pin down appropriate parameters for AdS/CFT models of the sQGP~\cite{Noronha:2009vz}.

\begin{figure}
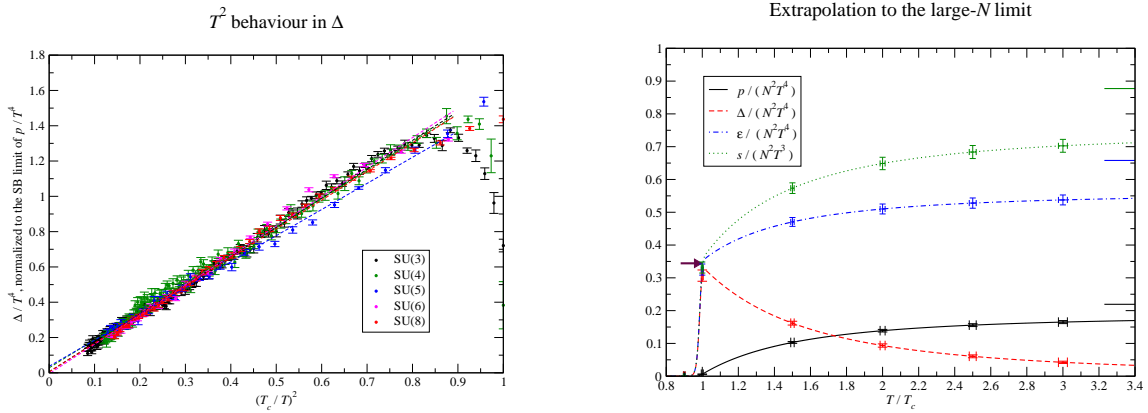

\centerline{\includegraphics[width=.44\textwidth]{Pisarski.eps} \hfill \includegraphics[width=.44\textwidth]{large_N_results.eps}}
\caption{Left panel: $\Delta/T^4$ ratio, plotted against $(T_c/T)^2$, appears to be compatible with the behavior described by eq.~(\protect\ref{T2_contributions_to_Delta}) in the range $(T_c/T)^2 \le 0.9$;  however, the precision of our data does not allow us to rule out the possibility that this quantity could also be described by more complicated functional forms, based on  combinations of perturbative logarithms. Right panel: extrapolation of $p/(N^2T^4)$ (black solid curve), $\Delta/(N^2T^4)$ (red dashed curve), $\epsilon/(N^2T^4)$ (blue dash-dotted curve) and $s/(N^2T^3)$ (green dotted curve) to the $N \rightarrow \infty$ limit; the errorbars (including statistical and systematic uncertainties) at some reference temperatures are also shown. The horizontal bars on the right-hand-side of this plot show the SB limits for the pressure, energy density and entropy density, from bottom to top; the maroon arrow denotes the large-$N$ limit of the latent heat $L_h$, as calculated in ref.~\protect\cite{Lucini:2005vg}.}
  \label{fig:Pisarski_and_large_N_results}
\end{figure}

Another issue we investigated is the possibility that the trace anomaly at temperatures close to $T_c$ may receive contributions proportional to $T^2$ of non-perturbative origin~\cite{T2_papers}. The simulations seem to confirm that this may be a general feature of all the gauge groups studied in this work: the left-hand-side panel of fig.~\ref{fig:Pisarski_and_large_N_results} shows that the results for $\Delta/T^4$ may be compatible with the behavior described by eq.~(\ref{T2_contributions_to_Delta}). However, we cannot rule out the possibility that the data may actually be described by a more complicated functional form (possibly involving logarithms of perturbative origin). Finally, the right-hand-side panel of fig.~\ref{fig:Pisarski_and_large_N_results} shows an extrapolation of our results to the $N \to \infty$ limit, based on the 
parametrization for the trace anomaly given in eq.~(C1) of ref.~\cite{Bazavov:2009zn}.

We plan to extend the present study of $\SU(N)$ thermodynamics to other observables which could be compared with gauge/gravity predictions~\cite{other_AdS_CFT_and_IHQCD_predictions}, and to renormalized Polyakov loops. On the other hand, to understand which non-perturbative features of the sQGP are directly related to the simplifications occurring in the large-$N$ limit and which are not, it would also be interesting to address similar high-precision thermodynamics studies in models in lower dimensions or based on smaller gauge groups, for which powerful numerical algorithms are available~\cite{smaller_gauge_group_works}.

\acknowledgments

The author acknowledges support from INFN, and thanks B.~Bringoltz, S.~Datta, Ph.~de~Forcrand, U.~G\"ursoy, E.~Kiritsis, I.~Kirsch, M.~Laine, M.~P.~Lombardo, F.~Nitti and G.~D.~Torrieri for correspondence and discussions. The author also thanks the authors of refs.~\cite{IHQCD_works} for providing the IHQCD curves reproduced in fig.~\ref{fig:comparison_with_IHQCD}.

\end{document}